# Lesion characterization using spectral mammography

B. Norell,[a] E. Fredenberg,[a,b] K. Leifland,[c] M. Lundqvist,[a] and B. Cederström[a,b]

[a]Philips Women's Healthcare, Smidesvägen 5, 171 41 Solna, Sweden;

[b]Department of Physics, Royal Institute of Technology (KTH), 106 91 Stockholm, Sweden;

[c]Department of Mammography, Capio S:t Görans Hospital, 112 81 Stockholm, Sweden;

## ABSTRACT

We present a novel method for characterizing mammographic findings using spectral imaging without the use of contrast agent. Within a statistical framework, suspicious findings are analyzed to determine if they are likely to be benign cystic lesions or malignant tissue. To evaluate the method, we have designed a phantom where combinations of different tissue types are realized by decomposition into the material bases aluminum and polyethylene. The results indicate that the lesion size limit for reliable characterization is below 10 mm diameter, when quantum noise is the only considered source of uncertainty. Furthermore, preliminary results using clinical images are encouraging, but allow no conclusions with significance.

## 1. INTRODUCTION

In breast-cancer screening, the most important task of the radiologist is early detection of breast cancer. The second most important task is to avoid recalling healthy women and exposing them to follow-up examinations, which may involve additional radiation or needle biopsy. Circular and oval lesions are, as opposed to stellate lesions, relatively easy to detect but difficult to characterize,[1] and most suspicious findings of this type are therefore called back and characterized using needle biopsy. A majority of these lesions are, however, benign cysts, and a method to characterize circular lesions as benign or malignant already at screening would have the potential to aid the radiologist in the latter task and reduce recall rates substantially.

Spectral imaging can be used to extract information about an object's constituents using the x-ray attenuation energy dependence, which is material specific.[2,3] The technology has been used in the past to improve on the signal-difference-to-noise ratio, to improve lesion conspicuity by reducing overlying anatomical structure, and to measure breast density.[4] We have developed a method that employs spectral imaging to characterize circular lesions as benign or malignant. Similarly to computer aided detection (CAD), the presented method can be used during screening examinations. The spectral mammogram is acquired as in conventional screening mammography; no additional time or dose is needed. In contrast to CAD, however, which primarily is used for detection and operates on the same image data as the radiologist, the method analyzes the constituents of a finding using information that is not available in a conventional mammogram.

The radiologist performs the analysis by marking a region of interest around the detected lesion as well as a reference region of normal tissue surrounding the lesion. Spectral information from the marked regions of interests is used to determine if the finding is likely to consist of water or cancerous tissue, with the underlying assumption that the contents of cystic benign lesions are similar to water.

Electronic mail: bjorn.norell@philips.com

## 2. MATERIAL AND METHODS

### 2.1. Description of the system

A spectral photon-counting mammography system has been developed based on the Philips MicroDose Mammography system. It is a scanning full-field digital mammography system with a tungsten-target x-ray tube that acquires high- and low-energy images, i.e. a spectral mammogram, with energy-sensitive silicon-strip detector units. All photons below a low-energy threshold at a few keV are rejected as noise. The remaining photons are separated by a high-energy threshold and contribute to either the low-energy or the high-energy image. A schematic of the system is shown in Fig. 1. Detailed descriptions of the system and detector can be found elsewhere.[4–10]

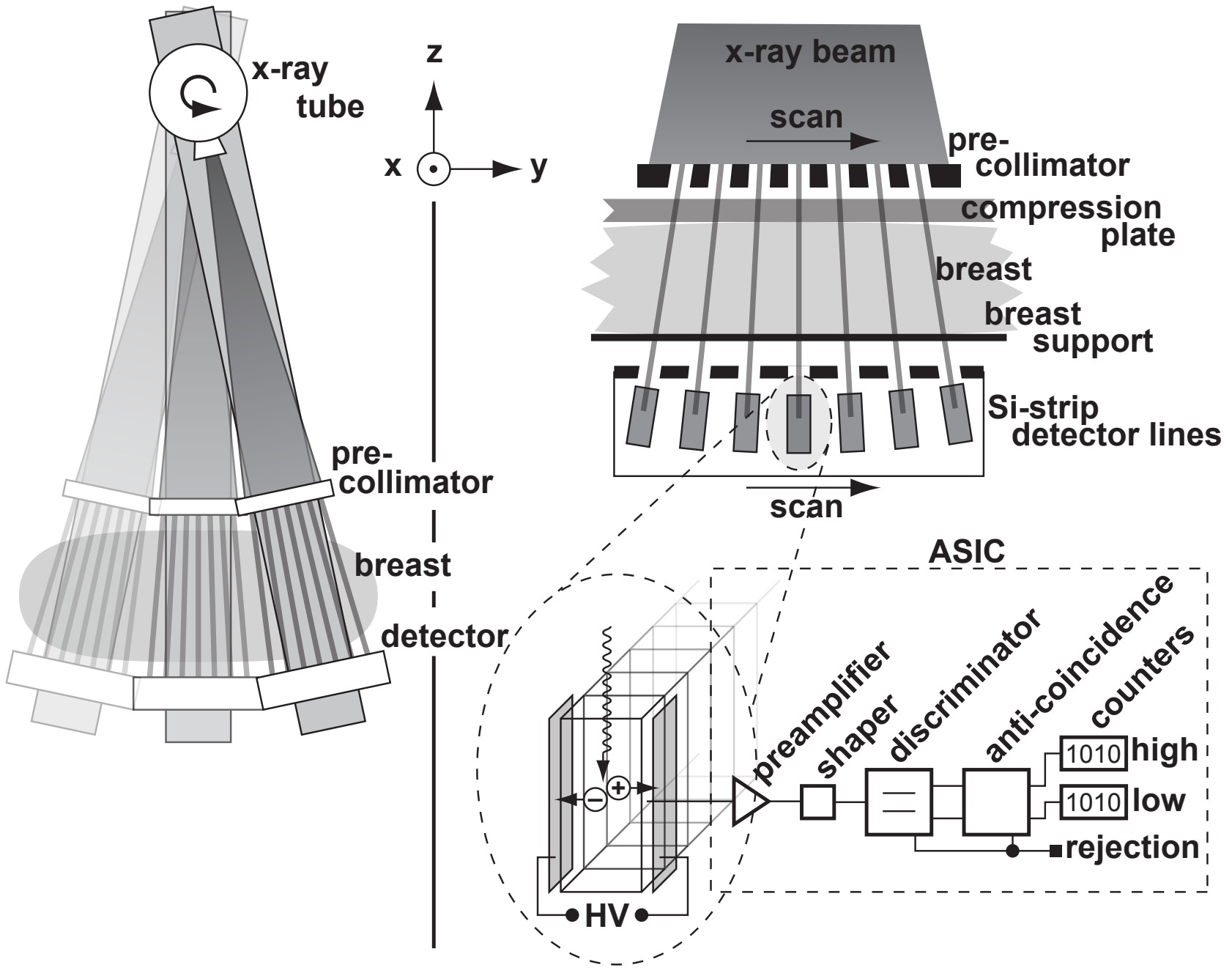

**Figure 1. Left:** Schematic of the Philips MicroDose Mammography system. **Right:** The image receptor and electronics.

The spectral mammogram is acquired with only one exposure and at no additional exposure time or radiation dose compared to a conventional mammogram. Nevertheless, a conventional mammogram is readily generated from the spectral data as the sum of the high- and low-energy images, and the system can therefore be used in regular screening. Whenever needed, the spectral information can be extracted and used for lesion characterization or other purposes.

The detected photon counts follow Poisson statistics, which makes statistical modeling fairly straightforward. However, a precise system model and careful calibration is required to predict system outcome.[4, 8–10]

### 2.2. Lesion characterization

Material decomposition between any two materials is achieved by solving the non-linear system of equations comprising high- and low-energy photon counts, the effective attenuation coefficients and the unknown thickness of each material. In the context of mammography, it is customary to assume a breast composition of adipose and fibro-glandular tissue and we solve the system of equations for the two unknown parameters breast thickness $h$ and breast glandularity $g$, i.e. the percentage of fibro-glandular tissue. A glandularity map and a thickness map are obtained by solving the system of equations for each image pixel.

However, when analyzing a suspicious lesion there are three unknowns for each pixel: breast thickness, breast glandularity, and lesion thickness, where the latter is a third material assumed to be either a malignant lesion or a cystic benign lesion. Therefore, a reference region of normal tissue around the detected lesion is used along with the region of interest that covers the lesion itself. We use this additional information to form a statistical model, i.e. the joint likelihood function, for whether the lesion is likely to be cystic (benign) or cancerous (malignant):

$$L(\theta|x,y) = L(h,g,t|x,y) = L(h,g|x)L(h,g,t|y), \tag{1}$$

where the model parameter vector $\theta$ can be expressed in terms of breast thickness $h$, glandularity $g$ and lesion thickness $t$. In other words, $\theta$ is a vector of input parameters to the system model that gives high- and low-energy photon counts as output. The observed data $x$ and $y$ are the high- and low-energy photons for the surrounding tissue and the lesion, respectively. These photon counts are Poisson distributed and are well approximated by the normal distribution $\mathcal{N}(n_i, \sqrt{n_i})$, and the likelihood $L$ for the statistical model with parameters $\theta$ given the observation $x$ is defined by the product

$$L(\theta|x) = \prod_k f(x_k|\theta), \quad \text{where} \quad f(x|n_i) \propto x \exp\left(-\frac{(x-n_i)^2}{2n_i}\right). \tag{2}$$

The joint likelihood is higher the closer the fit between the observed data and the statistical model. Equation (1) assumes that the breast thickness and glandularity is equal where the lesion is located as for surrounding tissue. The thickness $t$ is kept fixed to zero for the surrounding tissue, whereas $t$ for the lesion is a free variable.

Given the observed lesion and its surroundings, maximization of Eq. (1) for a cyst yields a figure of merit for the likelihood that the lesion is benign. The likelihood for a malignant tumor is computed in the same way. The two hypotheses are compared using the likelihood ratio

$$\lambda(x,y) = \frac{\max_{\theta_c} L(\theta_c|x,y)}{\max_{\theta_t} L(\theta_t|x,y)}, \tag{3}$$

where $\lambda$ is a confidence measure. For convenience, we study the log likelihood $\log(\lambda)$, i.e. for a likelihood ratio larger than one ($\log(\lambda) > 0$), benignity (cyst) is the more likely hypothesis. Conversely, the hypothesis of malignancy is more likely if $\log(\lambda) < 0$. A log likelihood ratio close to zero means that nothing can be said about malignancy or benignity.

In all calculations, normal breast tissue and malignant cancerous tissue were represented by published attenuation coefficients,[11] and benign cysts were assumed to contain pure water. It is clearly a simplification to assume circular lesions to be either solid tumors or cysts with fluid contents; benign cysts may be viscous rather than fluent, benign fibroadenomas are solid, and malignant mucinous carcinomas are viscous. These exceptions all appear with some frequency, but solid tumor versus liquid cyst remains a common and important differentiation problem.

### 2.3. Phantom design and measurements

A tissue-equivalent phantom comprising normal breast tissue, malignant lesions and cystic benign lesions was designed to experimentally evaluate spectral lesion characterization. Tissue equivalence was achieved by finding the combination of polyethylene and aluminum with the closest fit to the energy-dependent attenuation coefficients for the target material.[12] Published attenuation coefficients for normal and cancerous tissue was used,[11] and cysts were assumed to contain pure water. The lesions were 0, 10, 20, and 30 mm thick, embedded in normal breast tissue, and covered an area of $10 \times 10$ mm$^2$; see Fig. 2 for a schematic of the phantom. The design was validated by comparing the attenuation of simulated cysts (water) to compartments with actual water that were embedded in the phantom.

A flat-fielding algorithm was applied to each image but did not fully compensate for the heel effect, which resulted in a slight thickness gradient within $\pm 3\%$ perpendicular to the scan direction. To compensate for this effect, the low- and high-energy photon counts were rescaled with the same factor to get a thickness that was constant on average.

(a)

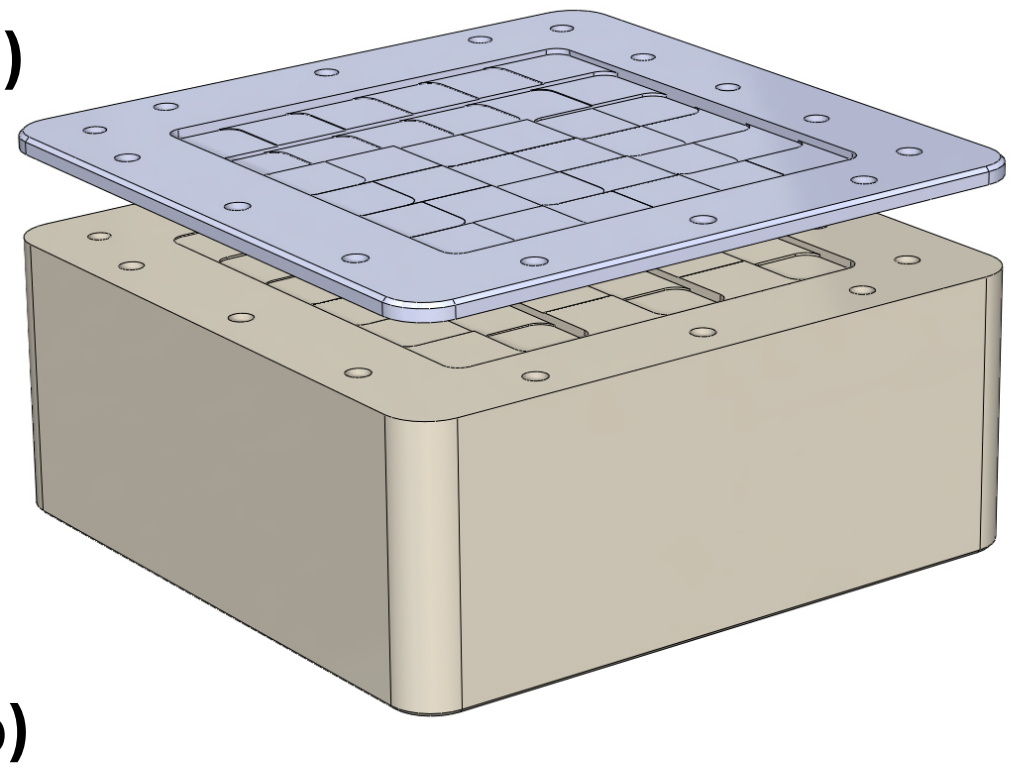

(b)

| | Material bases | |
|---|---|---|
| **Tissue** | **Polyethylene** | **Al** |
| glandular | 0.9811 | 0.0391 |
| adipose | 0.9637 | 0.0066 |
| skin | 1.0209 | 0.0402 |
| tumor | 0.9871 | 0.0429 |
| cyst | 0.9364 | 0.0416 |

(c)

| | | | | | | |
|---|---|---|---|---|---|---|
| gladularity range: | 100<br>4.925<br>0.196 | 90<br>4.918<br>0.181 | 80<br>4.910<br>0.167 | 70<br>4.902<br>0.152 | 60<br>4.894<br>0.137 | 50<br>4.886<br>0.123 |
| | 50<br>4.886<br>0.123 | 40<br>4.878<br>0.108 | 30<br>4.871<br>0.094 | 20<br>4.863<br>0.079 | 10<br>4.855<br>0.064 | 0<br>4.847<br>0.050 |
| lesion thickness: 0.5 cm | 4.859<br>0.068 | 4.833<br>0.067 | 4.894<br>0.133 | 4.868<br>0.132 | 4.928<br>0.198 | 4.903<br>0.197 |
| 1 cm | 4.871<br>0.086 | 4.820<br>0.085 | 4.901<br>0.143 | 4.850<br>0.142 | 4.931<br>0.200 | 4.881<br>0.198 |
| 2 cm | 4.894<br>0.122 | 4.792<br>0.120 | 4.916<br>0.163 | 4.814<br>0.160 | 4.937<br>0.204 | 4.836<br>0.201 |
| 3 cm | 4.917<br>0.159 | 4.765<br>0.155 | 4.930<br>0.183 | 4.778<br>0.179 | 4.943<br>0.207 | 4.791<br>0.204 |
| | tumors, 0% gland. | cysts, 0% gland. | tumors, 50% gland. | cysts, 50% gland. | tumors, 100% gland. | cysts, 100% gland. |

% gland.

poly-ethylene (cm)

Al (cm)

**Figure 2.** Phantom design. **(a)** Schematic of the tissue-equivalent phantom that consists of polyethylene and aluminum in different proportions. **(b)** Material basis decomposition data used for the phantom design. **(c)** Matrix showing phantom design.

## 2.4. Clinical pilot study

A pilot study to clinically evaluate spectral imaging was performed at Unilabs Mammography Department, Capio S:t Görans Hospital, Stockholm, Sweden in 2010. Women with suspicious findings were asked to take part in the study and undergo examination with the spectral mammography system. A total of 12 women were examined during the course of the study.

All lesions were indicated by experienced radiologists, and pathological examination provided ground truth. The spectral analysis according to above, including marking of lesion and surrounding tissue and calculation of the likelihood ratio, was performed by medical physicists.

The clinical study was conducted with an early prototype system that differed slightly from the system used for the more recent phantom measurements. In particular, the energy resolution of the prototype was coarser and it is expected that the results would improve with the current system.

# 3. RESULTS

## 3.1. Phantom measurements

The measured contrast between simulated water (i.e. decomposed into Al and polyethylene) and actual water acquired at a range of energy spectra (26–38 kV) was within 0.1%–1.7% for both low- and high-energy photon counts. For the ratio low-energy photons to high-energy photons, the difference between simulated and actual water was even smaller ($< 0.5\%$). The largest deviation was found for the largest lesion thickness (30 mm). The deviations were considered sufficiently small to trust the phantom results.

Figure 3 shows the low- and high-energy photon counts for 0–30-mm cysts and tumors embedded in a 50-mm breast with 50% glandularity. Expected counts (solid lines) are compared to phantom measurements (dots). The grid lines show the curvilinear coordinate system in terms of breast thickness and glandularity, computed using the direct mapping between photon counts and breast composition. The measurement uncertainties due to noise in terms of standard deviations are shown as error bars. Note that the $y$-axis in Fig. 3 shows a weighted difference between high- and low-energy photons rather than just the high-energy photons. This weighting was convenient to enhance differences and visualize the results, but was not used in any calculations.

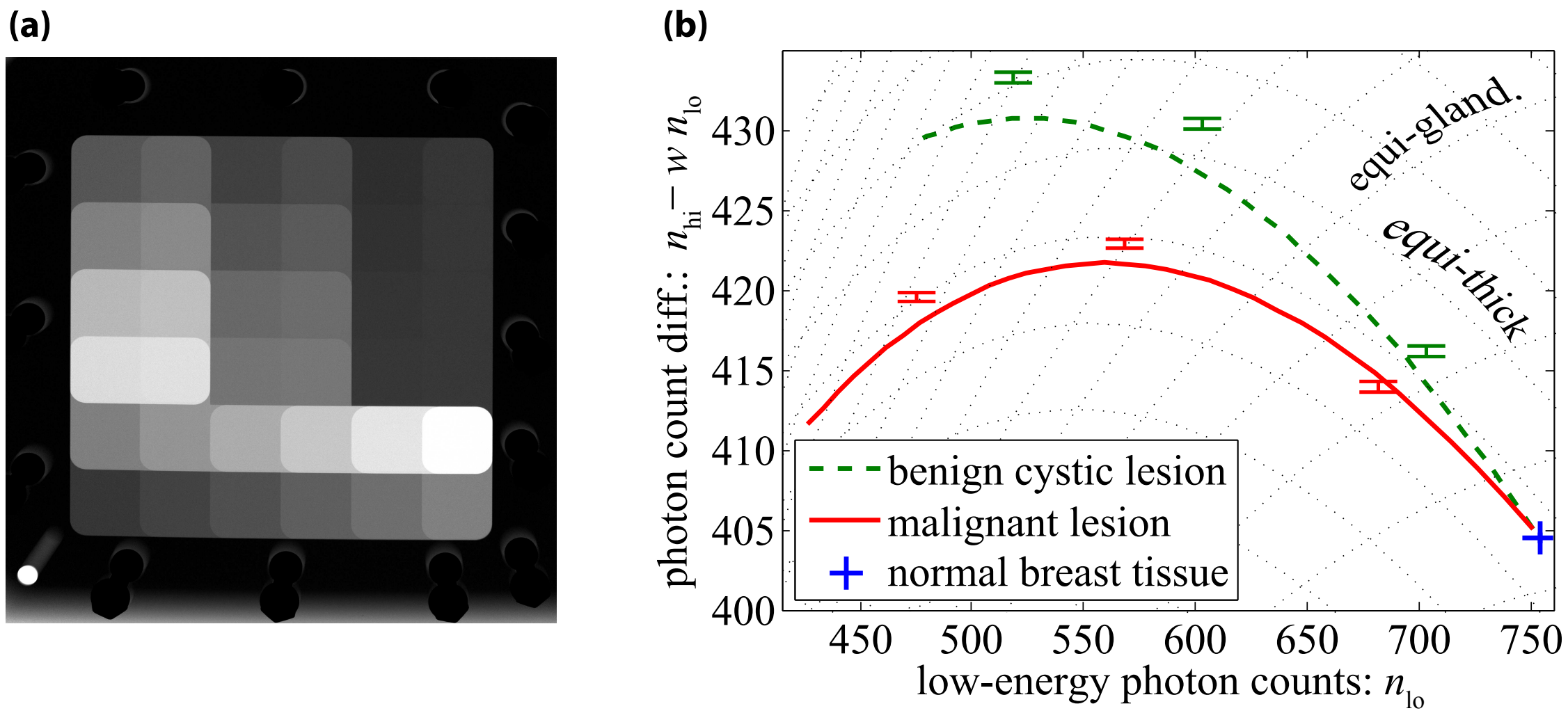

**Figure 3.** Phantom measurements. **(a)** X-ray image of the phantom. **(b)** Photon-count difference between cystic benign lesions (green, dashed) and malignant tumors (red, solid) as a function of low-energy counts. Expected counts (i.e. model predictions, lines) are compared to measurements on lesions (error bars) and the reference measurement (blue cross with error bar).

The phantom lesions were characterized at 29 kV and with the 0-mm lesion as reference. All lesions were correctly characterized with likelihood ratios $\log(\lambda) = -4, -100, -240$ for respectively 10-, 20-, and 30-mm tumors (red dots in the figure), and $\log(\lambda) = 3$, 54 and 140 for the cysts (green dots).

Quantum noise sets a lower limit on lesion size that can be accurately characterized. A log likelihood $|\log(\lambda)| = 4$ is equivalent to a separation of two standard deviations between the most likely malignant lesion and the most likely cystic lesion. For spherical lesions at clinical dose levels, this typically occurs at lesion diameters less than 10 mm. However, the situation improves for elliptical lesions if the area cross-sectional to the x-ray path is larger than the lesion thickness.

### 3.2. Clinical pilot study

Figure 4 shows the graphical user interface for the software tool and the results for a 13-mm benign cystic lesion. The best fit to the observed data that maximizes the likelihood function was obtained for a 6 mm thick cyst embedded in 39 mm breast tissue with 28% glandularity. A positive likelihood ratio of $\lambda = 30$ indicated that the lesion was likely to contain water and therefore probably benign. The compression height (40 mm in this case) serves as a sanity check, but is a too coarse measure to be trusted as ground truth.

Figure 5 illustrates calculation of the likelihood ratio for the case in Fig. 4. The 2D histogram in the left-hand panel shows the distribution of high- and low-energy counts over the marked lesion and reference area. In qualitative accordance with Fig. 3, the red and green solid lines in Fig. 5 represent model predictions of benign and malignant lesions with increasing thickness from 0 to $t$. The right-hand panel of Fig. 5 shows a closeup of the model predictions with the mean of the measurements indicated. This case is clearly characterized as a cyst.

In total, 9 circular lesions (7 benign and 2 malignant ones) were observed during the study. Out of these, 7 were correctly characterized with a decision boundary at $\log(\lambda) = 0$. One lesion was incorrectly characterized and the remaining one could not be characterized at all because of non-constant breast thickness. Moreover, of the correctly characterized lesions, two were very close to the decision boundary.

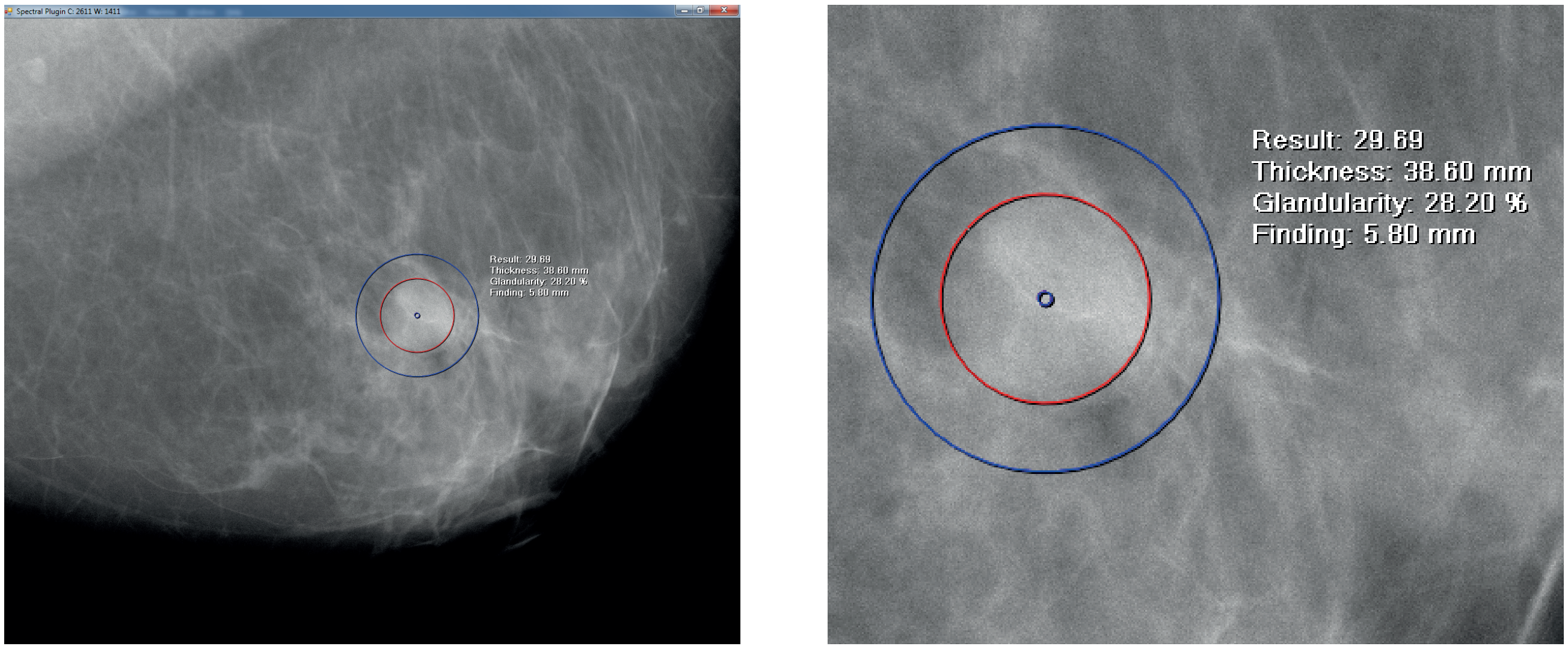

**Figure 4.** Overview (left) and closeup (right) of the user interface with a marked benign cystic lesion (red), the surrounding region that is used for reference (blue), and the resulting analysis: log likelihood, thickness, glandularity, and cyst thickness.

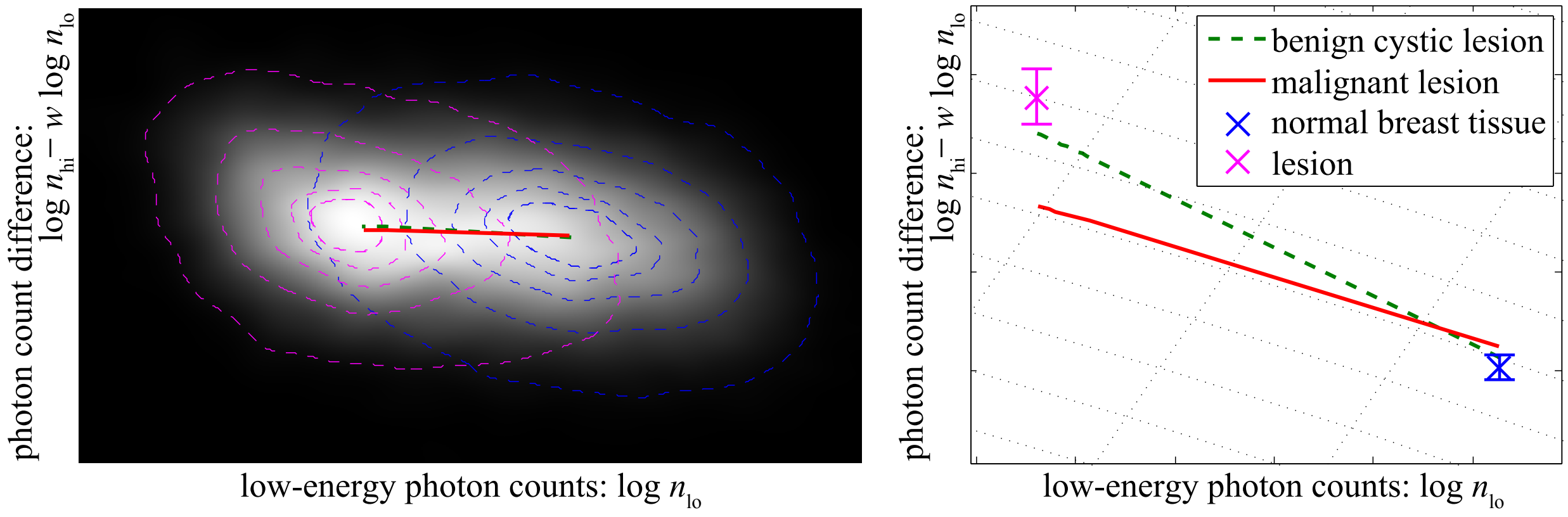

**Figure 5.** Photon-count difference between cystic benign lesions and malignant tumors as a function of low-energy counts. **Left:** Histograms over the marked lesion and reference area with superimposed solid lines that represent expected counts (model predictions). **Right:** Closeup of the expected counts compared to the mean of the histograms. [Color image available online.]

## 4. DISCUSSION

It must be emphasized that the likelihood function that was used is an approximation and does not include several systematic and statistical uncertainties. Therefore, the likelihood ratios ($\lambda$) given in Sec. 3.1 are expected to be reduced if additional uncertainties are included, and some preliminary findings are presented here. The assumption that both glandularity and thickness are equal for the lesion and surrounding region can be questioned. Fig. 3 shows that the gridlines for equal breast thickness are nearly parallel to the lines with varying lesion thickness. Thus, a difference in glandularity between the two regions should not affect the separation of cysts from solid tumors substantially, but only change the estimated lesion thickness.

A thickness difference between the lesion and background region, however, shifts a point almost perpendicularly to the lesion lines, thus seriously affecting the discrimination task. For a 1-cm lesion in a 5-cm breast, a gaussian uncertainty of the thickness difference of about a quarter of a mm, translates to half the separation between the cyst and solid tumor curves. The method is also sensitive to unknown inhomogeneities in the incoming x-ray field, such as a heel effect. We need to know the fluence ratio of the two regions to about 0.5% precision to have a two sigma separation.

A more detailed analysis of these uncertainties, as well as additional ones, such as threshold accuracy and stability, must be performed to fully evaluate the confidence of the lesion classification.

## 5. CONCLUSIONS

Measurements on a tissue-equivalent phantom showed that a significant difference between cysts with fluid contents and cancerous tissue can be detected with spectral lesion characterization at imaging conditions that are standard for screening. This result indicates that the specificity of mammographic screening could be increased. The amount of clinical data is too small to draw any final conclusions, but the results on clinical images are encouraging. Future work should include compensation for thickness gradients, accurate attenuation characterization of cyst fluid, and studies of the model accuracy.